\documentclass[12pt,a4paper,final]{iopart}
\usepackage{iopams}  
\usepackage{graphicx}
\usepackage[breaklinks=true,colorlinks=true,linkcolor=blue,urlcolor=blue,citecolor=blue]{hyperref}

\begin{document}

\title[Random Walker Models for Durotaxis]{Random Walker Models for Durotaxis}

\author{Charles R. Doering$^{1,2,3}$}
\address{$^1$Center for the Study of Complex Systems, University of Michigan, Ann Arbor, Michigan 48109-1107, USA}
\address{$^2$Department of Mathematics, University of Michigan, Ann Arbor, Michigan 48109-1043, USA}
\address{$^3$Department of Physics, University of Michigan, Ann Arbor, Michigan 48109-1040, USA}

\author{Xiaoming Mao}
\address{Department of Physics, University of Michigan, Ann Arbor, Michigan 48109-1040, USA}

\author[cor1]{Leonard M. Sander$^{1,2}$}
\address{$^1$Center for the Study of Complex Systems, University of Michigan, Ann Arbor, Michigan 48109-1107, USA}
\address{$^2$Department of Physics, University of Michigan, Ann Arbor, Michigan 48109-1040, USA}

\eads{\mailto{lsander@umich.edu}}

\begin{abstract}
Motile biological cells in tissue often display the phenomenon of \emph{durotaxis}, i.e. they tend to move towards stiffer parts of substrate tissue. The mechanism for this behavior is not completely understood. We consider  simplified models for durotaxis based on the classic persistent random walker scheme. We show that even a one-dimensional model of this type sheds interesting light on the classes of behavior  cells might exhibit. 
Our results strongly indicate that cells must be able to sense the \emph{gradient of stiffness} in order to show the effects observed in experiment. This is in contrast to the claims in recent publications that  it is sufficient for cells to be more persistent in their motion on stiff substrates to show durotaxis: i.e., if would be enough to sense the \emph{value} of the stiffness. We show that  these cases give rise to extremely inefficient transport towards stiff regions. Gradient sensing is almost certainly the selected behavior.

\end{abstract}

\noindent{\it Keywords}: Durotaxis, Persistent Random Walks

\section{Introduction}

In this paper we show that an extremely simple, schematic model of cell motility, the persistent random walker model, can give clues to aid in  the development of biologically realistic theories of durotaxis. 

Durotaxis, the tendency of cells to move towards stiff tissue was originally observed for fibroblasts moving on a surface with a sharp interface between stiff and compliant parts \cite{lo2000}. The cells tend to accumulate in the stiff region. Other experiments show that with a more gradual transition of stiffness, cells move up stiffness gradients \cite{vincent2013}. The biological function of this phenomenon is not completely clear, but it is known to be quite common \cite{kai2016}.  It may be a mechanism for cells to find a lesion \cite{vincent2013}. On the basis of this example  and others, it seems clear that durotaxis must have evolved to transport cells into stiff regions. This  remark will be important in our discussion below.

It is quite well known that cells have mechanisms to  measure the local stiffness of the matrix to which it is attached \cite{pelham1997}, basically  by contracting the attachments and sensing the response.  The attachments typically occur at many \emph{focal adhesions} where groups of proteins such as integrin traverse the cell membrane and dock on the substrate. The question we will address is how the cell uses the information from the different focal adhesions. 

On a coarse scale, cell motion may be thought of as a persistent  random walk; see  \cite{weiss2002} for a review. That is, the cell polarizes and moves a certain distance, then it turns by a greater or lesser angle, and continues in a new direction. In this picture, the path is governed by a few variables: the speed along the straight-line path, and the rate and angle of turning. Changes in these  parameters must depend on the information available to the cell. We will focus on the simplest kind of information which the cell is known to have (see above), the mechanical parameters of the substrate. There are two obvious candidates of this type:  the local stiffness, i.e. the average stiffness at the focal adhesions, or  differences in stiffness between the adhesions, i.e.,  the stiffness gradient.

Let us examine how durotaxis might arise  based on the first choice. One  might use the  known fact that cells move faster on stiff substrates \cite{discher2005,ulrich2009} and note that a cell entering a stiff region will run away, but that cells in the soft region will not. Thus there is net transport towards the stiff side. We will show, below, that this plausible-sounding reasoning is quite misleading. 

A closely related  idea notes  that cells have a slower rate of turning away from the current direction on stiff substrates than on soft ones \cite{missirlis2013}.  The path of the cell is more directed, and gives a higher average speed. The same argument as above is then invoked to say that there will be net transport towards the stiff side., and this has been the basis for theoretical modeling \cite{novikova2017,yu2017}. However, the argument has the same flaws as the previous one, as we will see.

Finally, we could postulate that the cell knows the gradient \cite{harland2011,feng2018}, and preferentially turns towards the up-gradient direction. We believe that any viable theory must be of this active-turning type. 

\section{Random Walkers with Persistence}
A stochastic scheme that represents persistent random walkers has been studied for a long time \cite{weiss2002}. In this paper we work with a one-dimensional version. Consider two populations of walkers,  one going in the up gradient direction, and the other down. The probability densities of being at position $x$ and time $t$ is $p(x,t)$ for the up-moving cells, and $q(x,t)$ for those moving down.  The walkers proceed with velocites $v_{\pm}$, which can depend on $x$. (For simplicity we assume that  the populations move with equal speeds in opposite directions, i.e. $v =v_+=-v_{-}$).
To represent changes in direction of the walkers we introduce two turning rates, $\gamma_{\pm}, \gamma_{\mp}$. These  are the rates of conversion $p \rightarrow q$, $q \rightarrow p$, respectively. We assume that the $\gamma$'s  can also be position-dependent.

\subsection{Master Equation}
We write the  master equation for the process is \cite{weiss2002} in terms of a discrete time-step, $dt$. We define  $dx = vdt$:
\begin{eqnarray}
\label{discretemaster}
p(x,t+dt) &=& (1- \Gamma_{\pm})  p(x-dx,t) + \Gamma_{\mp} q(x-dx,t) \nonumber \\
q(x, t+dt) &=& \Gamma_{\pm} p(x+dx,t) + (1- \Gamma_{\mp}) q(x+dx,t).
\end{eqnarray}
In this equation, $\Gamma_{\pm, \mp} = dt \ \gamma_{\pm,\mp}$. This equation can easily be directly simulated by random walker sampling. We present such results below.

It is also useful to compare to analytic results. To this end we consider $dt \to 0$. If we have the proper scaling of the parameters: that is if $\gamma_{\pm,\mp}$, $v$ are finite as $dt \to 0$,  we can write Eq.(\ref{discretemaster}) in the form of two continuity equations for the up-going and down-going currents:
\begin{eqnarray}
\label{continuum}
\partial_t p(x,t) + \partial_x(v p) &= & \gamma_{\mp} q(x,t) - \gamma_{\pm} p(x,t) \nonumber \\
\partial_t q(x,t) - \partial_x(v q) &= & -\gamma_{\mp} q(x,t) + \gamma_{\pm} p(x,t)
\end{eqnarray}

\subsection{Telegrapher's Equation}
In the  case when all parameters are independent of $x$, and $\gamma_{\pm}=\gamma_{\mp}=\gamma$ it is easy to use Eq. (\ref{continuum})  to derive an autonomous equation for the total probability $P=p+q$:
\begin{equation}
\label{telegraphers}
\partial_{tt} P + 2\gamma \ \partial_t P = v^2  \partial_{xx} P.
\end{equation}
This is the famous telegrapher's equation, first derived by Lord Kelvin \cite{thomson1856}. 

We will need to use qualitative features of the solutions to Eq. (\ref{telegraphers}) in what follows. One way to characterize solutions of equations such as Eq. (\ref{telegraphers}) is to consider moments of the distribution:
$$M_n = \int dx \ x^n P(x). $$
For example $M_0$ is the total probability (which needs to be conserved), $M_1$ gives the centroid of the distribution: its time dependence is the net drift, $M_2$ is the width, etc.

If we multiply Eq. (\ref{telegraphers}) by $x^2$ and solve for $M_2$. For initial conditions $M_2(0) = \partial_t M_2(0) =0$ we find:
  the well-known result:
 \begin{equation}
\label{2moment}
M_2 = v^2 \left( \frac{t}{\gamma} - \frac{1-\exp(-2\gamma t)}{2\gamma^2}  \right). 
\end{equation}
For short times, $t \ll 1/\gamma$, we have ballistic transport, $\left< x^2 \right>  \to v^2t^2$. For long times $t \gg 1/\gamma,\  \left< x^2 \right>  \to v^2t/\gamma$, i.e.  diffusive transport. In the long-time regime we can replace  Eq. (\ref{telegraphers})  with a diffusion equation with diffusion coefficient $v^2/2\gamma$.
 \subsection{Spatial Dependence}
Our interest here is in inhomogenous substrates. To model faster transport on stiff substrates, we take $v=v(x)$. Then, following the same path that led to Eq. (\ref{telegraphers}), we find:
\begin{equation}
\label{vofx}
\partial_{tt} P + 2\gamma \ \partial_t P =  \partial_{x} (v \ \partial_x (v P)).
\end{equation}
Note that this equation is consistent with conservation of $P$. This is easily seen by integrating over space. The right-hand side vanishes because it is a spatial derivative; thus solutions to Eq.(\ref{vofx}) can obey $\partial_t \int dx\ P \equiv \partial_t M_0 =0$.

 As far as we know, Eq. (\ref{vofx}) has not appeared in the literature. Several versions of the equation have appeared which do not conserve $M_0$, i.e. the number of random walkers is not constant. For example in \cite{masoliver1994,weiss2002} the right-hand side is $v \partial_x (v \partial_x P)$. This does not conserve probability, as is easily seen by integrating both sides over $x$.  Also in \cite{novikova2017} the rhs is $v^2 \partial_{xx} P$ which is also non-conserving. 
  
For the case of turning rate that depends on position, $\gamma = \gamma(x)$  \cite{novikova2017,yu2017} there is no closed form equation for $P$ \cite{weiss2002}. In this case we rely on simulations alone.  (In \cite{novikova2017} a version of Eq. (\ref{telegraphers}) is written with $\gamma(x) \partial_t P$ as the second term on the left, but this  equation does not represent any limit of Eq. (\ref{discretemaster}).) 

For a sharp interface on both sides of the boundary we will have Eq. (\ref{telegraphers}) with different values of $v, \gamma$. In particular, in the diffusion limit ($t \gg 1/\gamma$) we are allowed to consider  a diffusion equation  with different diffusion coefficients for $x>0$ and $x<0$. We give, below, an exact solution for this case.

Finally, we turn to the case of active turning. Suppose we are in a region with constant stiffness gradient. Durotaxis in this case means that the up-going walker will have a small rate  to turn around, and the down-going walker will have a large rate. We can write: $$\gamma_{\pm} = \gamma- \Delta, \  \gamma_{\mp} = \gamma +\Delta, \ \Delta > 0.  $$
If we insert these  relations into Eq. (\ref{continuum}) and form an equation for $P$ we find:
\begin{equation}
\label{drift}
\partial_{tt} P + 2\gamma\ \partial_t P = \partial_x(v \ \partial_x vP)  - 2\Delta \partial_x(vP).
\end{equation}
This is a generalization of the  telegrapher's equation to include diffusion and drift. In the long-time regime, for constant $v$, the diffusion coefficient is $v^2/2\gamma$ and the drift velocity is $v\Delta/\gamma$.

\section{Results and discussion}
We now give results for several situations to represent cells undergoing durotaxis. We give simulation results in all cases, and analytic results where available.  We always consider releasing a large number of non-interacting cells near $x=0$ and letting the distribution develop in time.

\subsection{Velocity increases with $x$}
We model a slowly increasing walker speed with the following formula:
\begin{equation}
\label{vX}
v(x)= v_o \frac{(x+X)}{X}.
\end{equation}
We take  $v_o=1$ and $X=1000$. 
In the analytical results that follow we consider cases where $$\epsilon=v_o/2\gamma X \ll 1, $$ 
which amounts to assuming that the gradient in stiffness is small.  

Analytical results for this case can be derived from Eq. (\ref{vofx}). We can take moments of the equation using Eq. (\ref{vX}). For  the first moment, which measures the drift of the distribution we get:
\begin{equation}
\label{M1X}
\frac{d^2 M_1}{dt^2} + 2\gamma \frac{dM_1}{dt} = v_o^2 \left( \frac{1}{X} + \frac{M_1}{X^2} \right).
\end{equation}
In the regime $1/\gamma < t < \gamma X^2/v_o^2$ we have the leading behavior:
\begin{equation}
\label{M1leading}
M_1 = \frac{v_o^2}{2\gamma X}t = \epsilon v_o t.
\end{equation}
A similar calculation shows that the leading term for $M_2$, the width of the distribution, is $(v_o^2/\gamma)t$. 

\begin{figure}[htbp]
\centering
\includegraphics[width=0.7\textwidth]{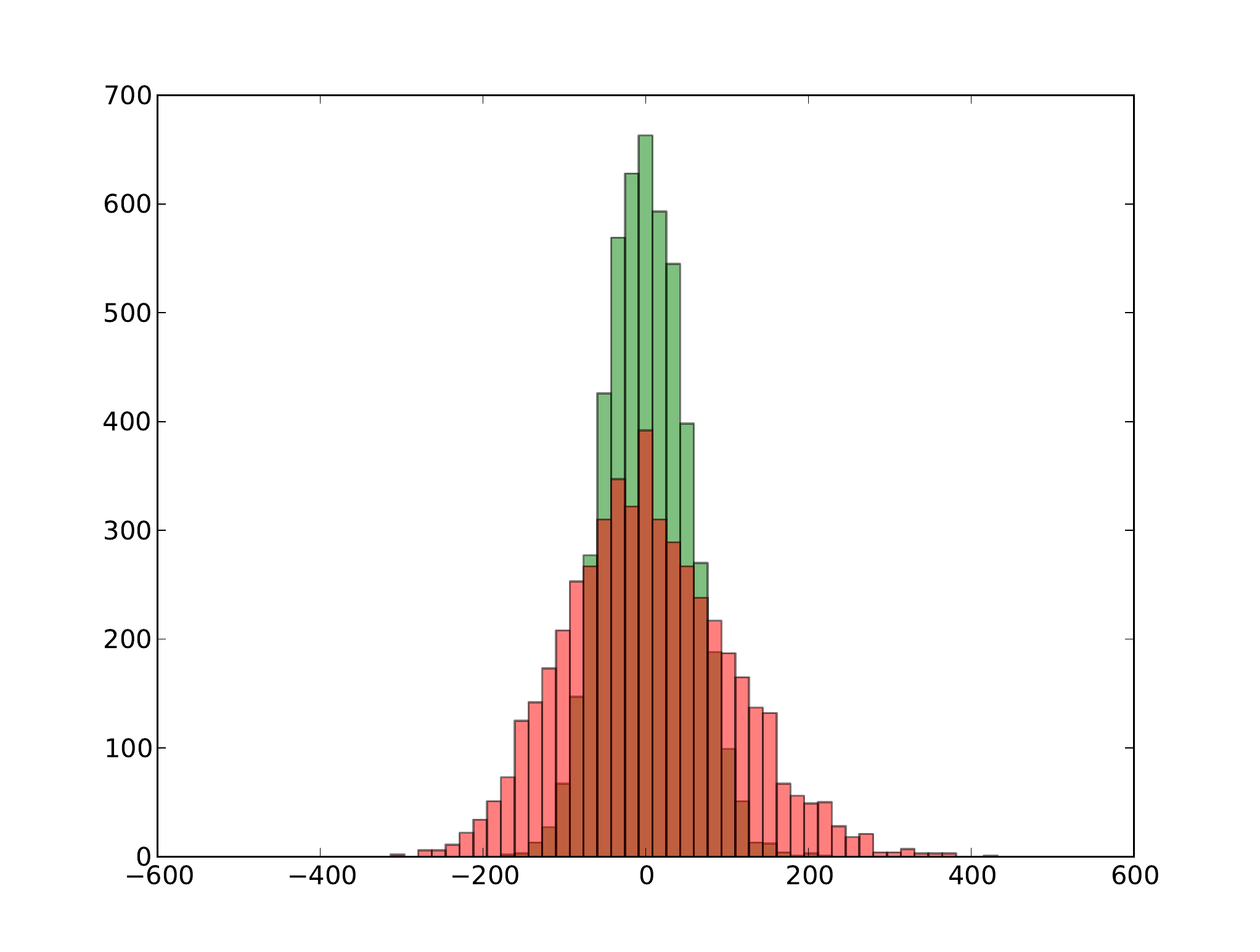} 
\caption{Histogram of the final positions of 5000 random walkers after 250 time steps (green) and 1000 time steps (red) with $v$ given by Eq. (\ref{vX}). $X=1000, v_o=1, \gamma =0.1.$}
 \label{figvofx}
\end{figure}

These results seem to be consistent with (slow) drift plus diffusion. However, simulations show that the moments alone are misleading. In Figure \ref{figvofx} we show a histogram of the distribution of walkers for a realization of the process. A measurement of the centroid of the distribution (not shown) agrees with Eq. (\ref{M1leading}). However,
note that because the walkers on the right (stiff) parts of the substrate move fast, and give rise to a long tail of the distribution. Those on the left are ``left behind". In all cases we have simulated many walkers remain at $x<0$. Thus having $v$ increase with $x$ is not an efficient way to transport a population up a stiffness gradient. Instead, it gives rise to a few fast walkers.

\subsection{Turning rate decreases with $x$}
\begin{figure}[htbp]
\centering
\includegraphics[width=0.7\textwidth]{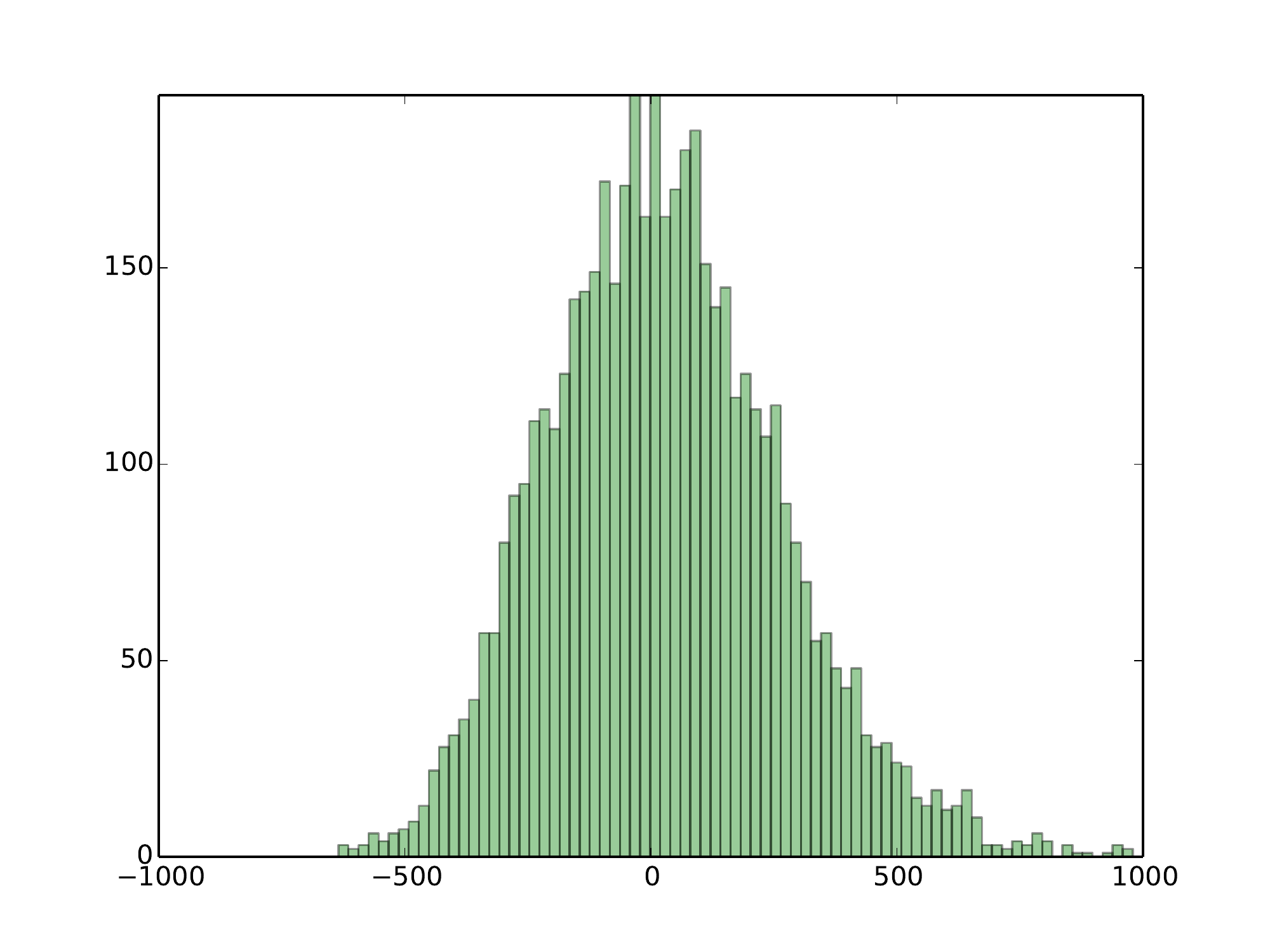} 
\caption{Histogram of the final positions of 5000 random walkers after 5000 time steps with $\gamma$ given by Eq. (\ref{gofx}). $X=1000, v_o=1.$}
 \label{figgofx}
\end{figure}

Recent theoretical modeling \cite{novikova2017,yu2017} has considered the case where persistence depends on stiffness. As we have argued above, this corresponds to $\gamma$ depending on $x$. We considered the specific case:
\begin{equation}
\gamma(x) = 0.1(X-x)/X,
\label{gofx}
\end{equation}
with $X=1000$ as before. In order to avoid negative turning rates we set a lower limit for $\gamma$ for large $x$, namely $\gamma = 0.01, x>0.9X$.
And for $x < -4X$ we set  $\gamma = 0.5$. A simulation of this case is given in Figure \ref{figgofx}. Once more we have the oddly skewed distribution with many of the walkers not transported to large $x$. Once more, as in the previous case, the centroid of the distribution increases approximately linearly in time as does the width. 

\subsection {Sharp interface with larger diffusion coefficient for $x>0$}
For a sharp interface, in the diffusion limit, $t \gg 1/\gamma$, the problem is exactly solvable using classical methods. This is due to the fact that for the diffusion limit  $D=v^2/2\gamma$ is the only important quantity. If either $v$ or $\gamma$ (or both) differ on the sides of the interface but are constants, we have  to solve: 
\begin{eqnarray}
\partial_t P &=& D_+ \partial_{xx}P, \quad x>0, \nonumber \\
\partial_t P &=& D_- \partial_{xx}P, \quad x<0, 
\end{eqnarray}
where $D_+ = v_+^2/2\gamma_+, D_-=v_-^2/2\gamma_-$.
The initial condition is 
\begin{equation}
\label{initial}
P(x,t=0) = \delta(x-x_o),
\end{equation}
 for some small positive $x_o$. 
At the interface we need continuity of $P$ and of flux:
\begin{eqnarray}
\label{continuity}
P(0^+,t) &=& P(0^-,t), \nonumber \\
D_+\partial_x P(0^+,t) &=& D_- \partial_{x}P(0^-,t), 
\end{eqnarray}

With the initial condition of Eq. (\ref{initial}) we are solving for the Green's function for this problem. 
For the case $D_+=D_-$ the Green's function is well known:
\begin{equation}
\label{freegreen}
G(D,x-x_o) = \frac{\exp[-(x-x_o)^2/4Dt]}{\sqrt{4\pi Dt}}
\end{equation}

\begin{figure}[htbp]
\centering
\includegraphics[width=0.7\textwidth]{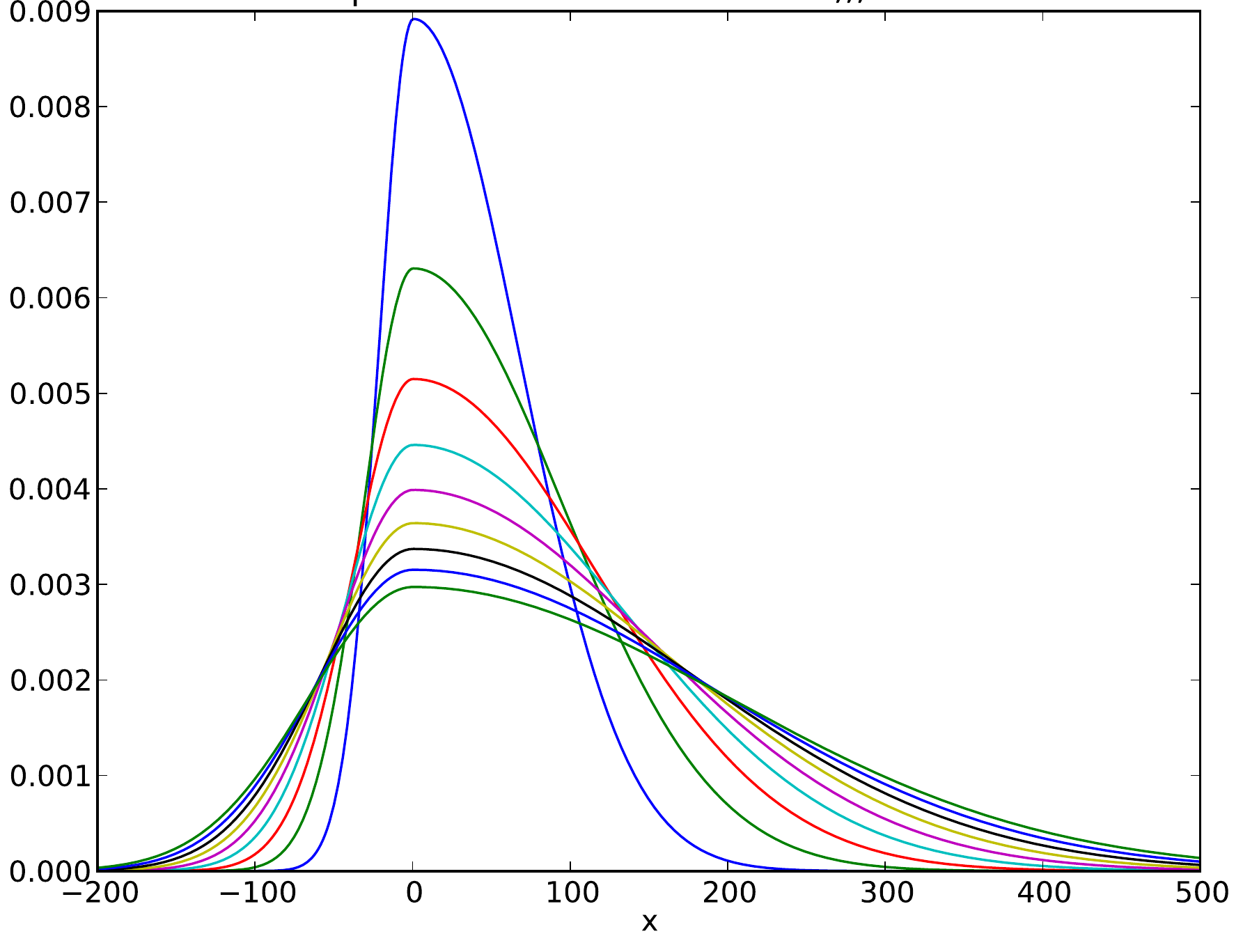} 
\caption{Plot of Eq. (\ref{twosidegreen}) for $D_+=2.25, D_-=0.25$ and $t= 1000 ... 10000$.}
 \label{figtwoside}
\end{figure}

We can solve the two-sided problem by an extension of the method of images \cite{jackson1999}, that is by superposing functions of the form of Eq. (\ref{freegreen}). The steps are straightforward and follow the corresponding problem in electrostatics. The solution is as follows: set 
$$ k=\sqrt{D_-/D_+}, \quad x_1 = k x_o. $$
Then the Green's function for the two-sided problem is:
\begin{eqnarray}
\label{twosidegreen}
G(D_+,x-x_o) + \frac{1-k}{1+k} \ G(D_+,x+x_o) & & \quad x>0 \nonumber \\
\frac{2k}{1+k} \ G(D_-, x-x_1) & & \quad x<0.
\end{eqnarray}
It is clear that the first line does satisfy the diffusion equation for $x>0$ and the second line for $x<0$. A bit of algebra demonstrates that they obey the continuity conditions of Eq. (\ref{continuity}). This solution is slightly more complicated than the electrostatics problem in that $x_1 \ne x_o$. This change is necessary so that the continuity conditions are satisfied for all times. 

In Figure \ref{figtwoside} we show the solution for various times for the case $k=1/3$. Once more we have a skewed distribution, qualitatively similar to the previous cases. Again, transport towards positive $x$ is not achieved except for a few fast cells.

\subsection {Active turning in a constant stiffness gradient}
\begin{figure}[htbp]
\centering
\includegraphics[width=0.7\textwidth]{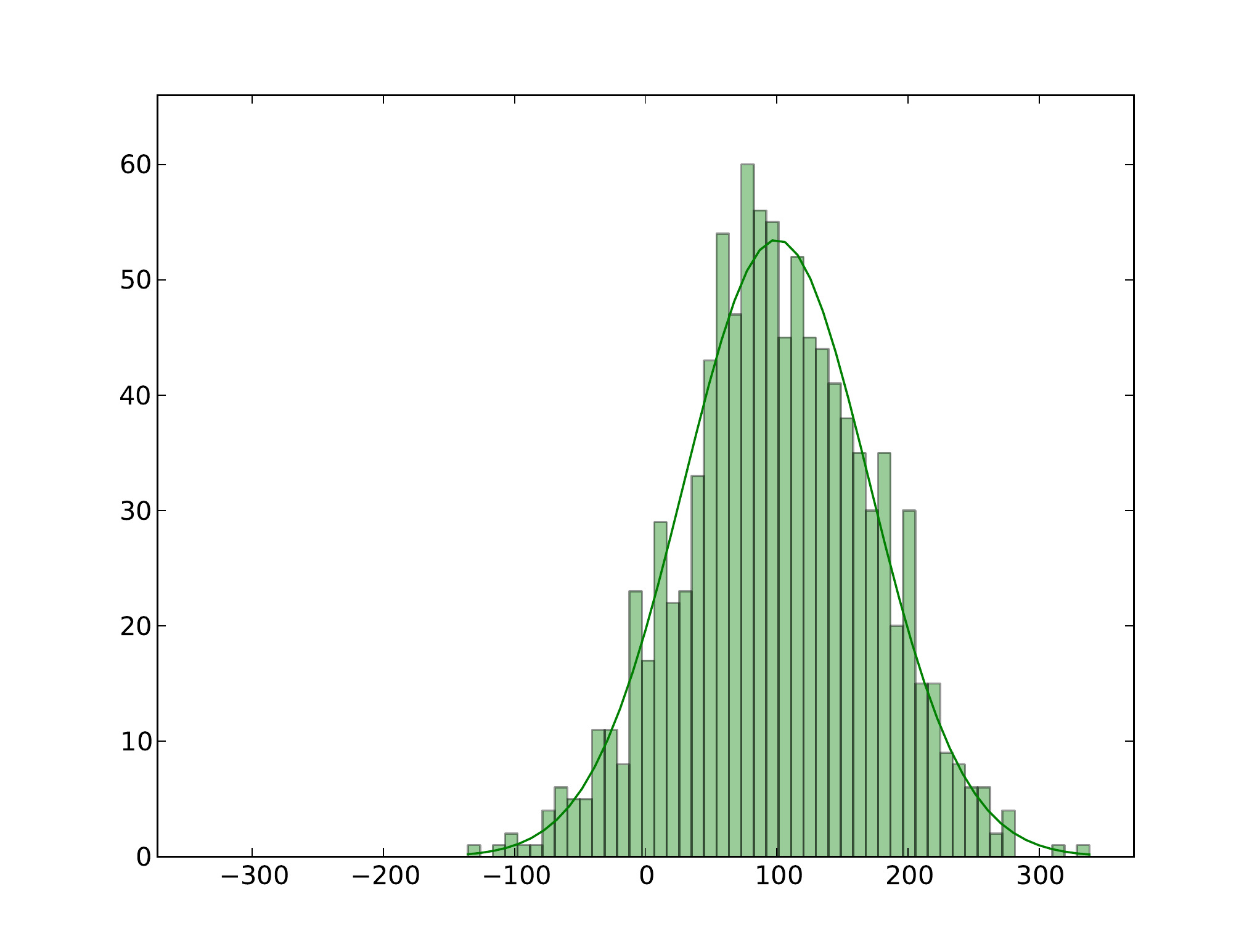} 
\caption{Simulation of 1000 random walkers for $t=500$ with $\gamma_+=0.08, \gamma_-=0.12$. The line is from Eq. (\ref{driftdiff}).  }
 \label{figdrift}
\end{figure} 

If there is active turning we have argued that $\gamma_\pm < \gamma_\mp$. We can simulate this case: the results are shown in Figure \ref{figdrift}.
Note that the distribution moves as a whole towards the right. Of course, in this case we can easily solve the continuum equation, Eq. (\ref{drift}) for the case $t \gg 1/\gamma$. The well-known result is:
\begin{equation}
\label{driftdiff}
P(x,t)=\frac{\exp[-(x-V t)^2/4Dt]}{\sqrt{4\pi Dt}},
\end{equation}
with $D=v^2/2\gamma$ and  $V=v\Delta/\gamma$ as we have remarked above. This result is also shown in Figure \ref{figdrift}.

This is the behavior that we expect to be relevant to real biology. There is efficient transport of cells up the stiffness gradient.


\section{Summary and Conclusions}
The reason we introduced a brutally simplified model in this paper is that  models of durotaxis which include such biologically relevant details such as the distribution of  focal adhesion and the details of cell contraction can be very complicated indeed \cite{harland2011,feng2018}. It is often very difficult to see the behavior of even a few cells for a short time, and studies of large populations such as those we have done are not practical. However, if we assume that durotaxis has a biological function, it is precisely the behavior of populations of cells that are relevant, and selected. We believe that our simplified model captures the essential physics in this regime. 

Of course, for motion on a substrate models should be two-dimensional. We have done simple preliminary simulations of a persistent walker model on a square lattice which shows identical effects to those we have presented.

Our conclusion from the simple model is clear: merely having faster speeds on stiffer substrates or more persistence on stiffer substrates will not efficiently move cells towards stiff regions. We think that the most promising way to model durotaxis is to formulate models that directly sense the stiffness gradient. 

Experiments can, in principle, show whether cells are measuring gradients of stiffness, and, indeed, many such experiments are in the literature. Typically they show that details of motility are affected only by gradients \cite{vincent2013}. Of course, cells on a substrate are a complicated system, and dependence on the absolute value of the stiffness cannot be ruled out.

\section*{Acknowledgements}
We have had very useful conversations on this subject with H. Levine and J-C Feng. CRD is supported by NSF- DMS-1515161 and XM by NSF-DMR-1609051.

\bibliographystyle{unsrt} 
\end{document}